\documentclass[preprint,authoryear,12pt]{elsarticle}
\usepackage{lineno}
\usepackage[utf8]{inputenc}
\usepackage{amsmath}
\usepackage{amsfonts}
\usepackage{amssymb}
\usepackage{makeidx}
\usepackage{enumitem}
\usepackage{mdframed}
\usepackage{natbib}
\usepackage[T1]{fontenc}
\usepackage{amsmath}
\usepackage{amssymb}
\usepackage{epsfig}
\usepackage{graphics}
\usepackage{dsfont}
\usepackage{natbib}
\usepackage{inputenc}
\usepackage{float}
\usepackage{hyperref}
\usepackage{fix-cm}
\usepackage{makeidx}
\usepackage{color}
\usepackage{url}
\usepackage{amsmath}
\usepackage{amsfonts}
\usepackage{amssymb}
\usepackage{makeidx}
\hypersetup{
    colorlinks=true, 
    linktoc=all,     
    linkcolor=blue,  
    citecolor=blue,
    filecolor=blue,
    urlcolor=blue,
}

\usepackage{framed}
\usepackage{soul}
\usepackage{flafter}
\usepackage{tablefootnote}
\usepackage{ftnxtra}
\usepackage{lipsum}
\usepackage{float}
\usepackage{arydshln}
\usepackage{hhline}
\floatstyle{plain} 
\restylefloat{table}

\usepackage{dblfloatfix}

\journal{Computational Statistics and Data Analysis}

\newcommand{\logit}{\text{logit}}
\newcommand{\llog}{\text{log-log}}

\begin{document}

\begin{frontmatter}

\title{A new Bayesian regression model for counts in medicine}

\author[label1]{Hamed Haselimashhadi}
\author[label1]{Veronica Vinciotti\corref{cor1}}
\author[label1]{Keming Yu}
\cortext[cor1]{Department of Mathematics, 208 John Crank Building, UB83PH Middlesex, UK; veronica.vinciotti@brunel.ac.uk; tel: +44 (0)1895267469}
\address[label1]{Department of Mathematics, Brunel University London, UB8 3PH Uxbridge, UK}




\begin{abstract}
Discrete data are collected in many application areas and are often characterised by highly skewed and power-lawlike distributions. An example of this, which is considered in this paper, is the number of visits to a specialist, often taken as a measure of demand in healthcare. A discrete Weibull regression model was recently proposed for regression problems with a discrete response and it was shown to possess two important features: the ability to capture over and under-dispersion simultaneously and a closed-form analytical expression of the quantiles of the conditional distribution. In this paper, we propose the first Bayesian implementation of a discrete Weibull regression model. The implementation considers a novel parameterization, where both parameters of the discrete Weibull distribution can be made dependent on the predictors. In addition, prior distributions can be imposed that encourage parameter shrinkage and that lead to variable selection. As with Bayesian procedures, the full posterior distribution of the parameters is returned, from which credible intervals can be readily calculated. A simulation study and the analysis of four real datasets of medical records show promises for the wide applicability of this approach to the analysis of count data. The method is implemented in the R package {\tt BDWreg}.
\end{abstract}


\end{frontmatter}


\section{Introduction}

Data in the form of counts appear in many application areas, from medicine, social and natural sciences to econometrics, finance and industry \citep{cameron2013regression}. In medicine, two examples of this are the length of stay in hospital, commonly used as an indicator of the quality of care and planning capacity within a hospital \citep{atienza08,carter14},  and the number of visits to a specialist \citep{machado05}, often taken as a measure of demand in healthcare. Other examples are high-throughput genomic data generated by next generation sequencing experiments \citep{ozsolak11, bao14, robinson08} or lifetime data, such as the number of cycles before a machine breaks down \citep{nakagawa75}.

Similarly to Weibull regression, which is widely used in lifetime data analysis and survival analysis for continuous response variables, \cite{kalktawi15} have recently proposed a regression model for a discrete response based on the discrete Weibull distribution. A number of studies have found a good fit of this distribution in comparison with other distributions for count data \citep{bracquemond03,englehardt11,lai13}. In the context of regression, \cite{kalktawi15} show two important features of a discrete Weibull distribution that make this a valuable alternative to the more traditional Poisson and Negative Binomial distributions and their extensions, such as Poisson mixtures \citep{hougaard97}, Poisson-Tweedie \citep{esnaola13}, zero-inflated regression\citep{lam06} and COMPoisson \citep{sellers10}: the ability to capture over and under-dispersion simultaneously and a closed-form analytical expression of the quantiles of the conditional distribution.

In \cite{kalktawi15}, maximum likelihood is used for the estimation of the parameters. This is in general the most common approach for parameter estimation in regression analysis of counts, due to a lack of simple and efficient algorithms for posterior computation \citep{zhou12}. Among the contributions to Bayesian estimation of discrete regression models, \cite{el1973bayesian} consider the case of Poisson regression, \cite{zhou12} provide an efficient Bayesian implementation of negative Binomial regression, \cite{mohebbi2014disease} develop Bayesian estimation for a Poisson and negative Binomial regression with a conditional autoregressive correlation structure whereas \cite{angers2003bayesian,ghosh2006bayesian,neelon2010bayesian,liu2012bayesian} study zero-inflated Poisson regression. In this paper, we contribute to this literature, by providing the first Bayesian approach for parameter estimation in discrete Weibull regression. For the choice of prior distributions, we consider both the case of non-informative priors and the case of Laplace priors with a hyper penalty parameter. The choice of Laplace priors induces parameter shrinkage \citep{park2008bayesian,kyung2010penalized}, and, with the use of Bayesian credible intervals, leads to variable selection, similar to alternative approaches such as spike and slab priors \citep{ishwaran2005spike}.

The aim of this paper is two-fold. Firstly, we highlight the role that the discrete Weibull distribution has in modelling count data from a variety of applications, beyond its current limited use to lifetime data. We particularly emphasize applications in the medical domain, using several  datasets of medical records. Secondly,  we present a novel Bayesian regression model for counts based on the assumption of a discrete Weibull conditional distribution. The remainder of this paper is organized as follows. Section (\ref{sec:regression}) describes the discrete Weibull regression model, with a more general parametrization as that presented in \citep{kalktawi15}. Section (\ref{dw-model-fitting}) describes Bayesian parameter estimation for a discrete Weibull regression model. Section (\ref{DW:simulation}) presents an extensive simulation study, whereas Section (\ref{empirical data in medicine}) shows the analysis of real data via Bayesian discrete regression model and a comparison with existing approaches. Finally, we draw some conclusions in Section (\ref{Conclusion and discussion}).

\section{Discrete Weibull regression}\label{sec:regression}

\subsection{Discrete Weibull distribution}

The discrete Weibull distribution was introduced by  \cite{nakagawa75}, as a discretized form of a continuous Weibull distribution, similarly to the geometric distribution, which is the discretized form of the exponential distribution, and the negative Binomial, which is the discrete alternative of a Gamma distribution. In some papers, this is referred to as a type I discrete Weibull, as two other distributions were subsequently defined. \cite{bracquemond03} review the three different distributions and point out the advantages of using the type I distribution: it has an unbounded support, differently to the type II distribution, and it has a more straightforward interpretation, differently to the type III distribution.

If a random variable $Y$ follows a (type I) discrete Weibull distribution, then  the cumulative distribution function of $Y$ is given by
\[F(y; q, \beta)=  \left\{
  \begin{array}{ll}
     1-q^{(y+1)^{\beta}} & \quad \textrm{if} \,\, y = 0, 1, 2, \ldots \mbox{(jump points)}\\
     0 & \quad \textrm{if} \,\, y < 0
   \end{array} \right.\]
with $0<q<1$ and $\beta>0$ the shape parameters.
A similar definition can be given on the support $1,2,\ldots$. In this case, $F(y; q, \beta)=1-q^{y^\beta}$, for $y=1,2,\ldots$. Comparing this cdf with that of a continuous Weibull distribution with parameters $\alpha$ and $\gamma$, one can see that there is a direct correspondence between $\beta$ and $\gamma$, whereas $q$ in the discrete case corresponds to $\exp(-\alpha)$ in the continuous case \citep{khan89}.

Given the form of the cumulative distribution function, the discrete Weibull distribution has the following probability mass function:
\[p(y; q,\beta)=q^{y^{\beta}}-q^{(y+1)^{\beta}},\,\,\  y=0, 1, 2, ...\]
with $q$ and $\beta$ denoting the shape parameters. Throughout the paper, we will refer to this distribution as DW$(q,\beta)$.

\subsection{Inference for Discrete Weibull: Existing Approaches}

\cite{khan89} derive estimators of the parameters  $q$ and $\beta$ using the method of moments and a new method which they call the method of proportions, and they find a good performance for the latter.
Let $Y_1,\ldots,Y_n$ be a random sample from a DW$(q,\beta)$ distribution and denote $Z=\sum_{i=1}^n I(Y_i=0)$ and $U=\sum_{i=1}^n I(Y_i=1)$. Using the method of proportions, the following estimators of $q$ and $\beta$ are proposed:
\begin{eqnarray*}
\hat q &=& 1-\dfrac{Z}{n}\\
\hat \beta &=& \ln \Big[\ln\Big(1-\dfrac{Z}{n}-\dfrac{U}{n}\Big)/\ln\Big(1-\dfrac{Z}{n}\Big)\Big]/\ln(2).
\end{eqnarray*}
These estimators use only the zeros and ones in the sample. \cite{santos13} derive an improved estimator of $\beta$, by taking all observations into account. In particular, let $d_m$ be the maximum observed value of $Y$ and let $k=d_m-1$. If $d_m>2$, then the following improved estimator is proposed:
\[\hat \beta = \dfrac{1}{k}\sum_{d=1}^k\ln \Big[\ln\Big(1-\hat{F}(d)\Big)/\ln(\hat q)\Big]/\ln(d+1),\]
where $\hat{F}$ denotes the empirical cdf. When $d_m=2$, this estimator is equivalent to the one from \cite{khan89}. Note that in both cases, no estimates of $\beta$ can be obtained when $\hat q =1$, i.e. there are no zero counts in the observed data, or $\hat q =0$, i.e. all counts are zero. However, in other cases, the estimators perform relatively well, particularly in the case of small sample sizes, as we have checked with a simulation study (not shown here).

\cite{kulasekera94} considers maximum likelihood for the estimation of $q$ and $\beta$. The likelihood function for a discrete Weibull sample is given by:
\[L(q, \beta) = \prod_{i=1}^n \Big(q^{y_i^{\beta}}-q^{(y_i+1)^{\beta}}\Big),\]
the maximum of which can be found numerically.

There is no explicit work in the literature for building confidence intervals for discrete Weibull parameters, although standard asymptotic likelihood and bootstrap approaches can be used. The Bayesian approach that we devise in this paper will lead naturally to credible intervals for the parameters.

\subsection{Regression via a discrete Weibull}

Let $Y$ be the response variable with possible values $0,1, \ldots$, and let $X_1,\ldots,X_{p}$ be $p$ covariates.
We assume that the conditional distribution of $Y$ given $X$ follows a DW distribution with parameters $q$ and $\beta$. There are a number of possible choices to link the parameters $q$ and $\beta$ to the  covariates. In particular, we propose the following link functions:
\begin{enumerate}
	\item $q$ dependent on $X$ via
	\begin{align*} 
    & \log(-\log(q))=X \boldsymbol{\theta} \:\: \mbox{or}\\
	 & \log\Big(\frac{q}{1-q}\Big)=X \boldsymbol{\theta},\\
	\end{align*}
	where  $X=(1 \: X_1 \ldots X_{p})$ and $\boldsymbol{\theta}=(\theta_0 \ldots \theta_{p})'$.
	\item $\beta$ dependent on $X$ via
	\begin{align*} 
	\log(\beta)= X\boldsymbol{\gamma},
	\end{align*}
	where $\boldsymbol{\gamma}=(\gamma_0 \: \gamma_1 \ldots \gamma_{p})'$.
\end{enumerate}
The first parametrization was proposed by \citep{kalktawi15}, in line with the link function used in continuous Weibull regression. In this paper, we consider one additional parametrization for $q$ via a logit link function, which has proved to be rather effective for statistical inference, and a link also between the second parameter $\beta$ and the covariates, in order to capture more complex dependencies.

\section{Bayesian inference for discrete Weibull regression}\label{dw-model-fitting}

In this section, we discuss Bayesian estimation of the regression parameters $\boldsymbol{\theta}=(\theta_0\ldots\theta_{p})'$ and $\boldsymbol{\gamma}=(\gamma_0\ldots\gamma_{p})'$. The advantage of choosing Bayesian approaches over classical maximum likelihood inference is two-fold. Firstly, the possibility of taking  prior information into account and, secondly, the procedure returns automatically credible intervals for all parameters.

Given $n$ observations $y_i$ and $(x_{i1},\ldots,x_{i{p}})$, $i=1,\ldots,n$, for  the response $Y$ and the covariates $X$, respectively, and letting $x_i$  be the row vector $x_i=(1,x_{i1},\ldots,x_{ip})$, the likelihood for the most general case is given by
\begin{equation*}
	l(\boldsymbol{\theta},\boldsymbol{\gamma}|X,Y) = \prod_{i=1}^n
	 \bigg( (\frac{e^{x_i\boldsymbol{\theta}}}{1+e^{x_i\boldsymbol{\theta}}})^{{y}^{x_i\boldsymbol{\gamma}}}
	-
	 (\frac{e^{x_i\boldsymbol{\theta}}}{1+e^{x_i\boldsymbol{\theta}}})^{{(y+1)}^{x_i\boldsymbol{\gamma}}} \bigg).
\end{equation*}

We consider different prior distributions on $\boldsymbol{\theta}$ and $\boldsymbol{\gamma}$. Unfortunately, in the context of discrete Weibull regression, there are no conjugate priors. However, we will show in the simulation study how an uninformative prior achieves an acceptable rate of mixing as well as  comparable estimation to maximum likelihood. In addition, we consider a prior on the regression coefficients that encourages sparsity. In particular, we consider a Laplace prior for $\theta$ and $\gamma$, of the form
\begin{align*}
p(\boldsymbol{\theta}|\lambda) & =\frac{\lambda}{2}e^{-\lambda|\boldsymbol{\theta}|},\qquad \lambda>0,\\
p(\boldsymbol{\gamma}|\tau) & =\frac{\tau}{2}e^{-\tau|\boldsymbol{\gamma}|},\qquad \tau >0.
\end{align*}
For a given choice of $\lambda$ and $\tau$, maximising the posterior probability under these priors corresponds to maximising the $L_1$ penalised log-likelihood
\begin{align*}
\log l(\boldsymbol{\theta},\boldsymbol{\gamma}|X,Y) -\lambda\sum_{j=1}^{p} |\theta_j|-\tau\sum_{k=1}^{p} |\gamma_k|,
\end{align*}
as in the traditional lasso approach \citep{park2008bayesian,tibshirani1996regression}.
We further assume a Gamma(a,b) hyper prior for both $\lambda$ and $\tau$, leading to the posterior distribution
\begin{align*}
p(\boldsymbol{\theta},\boldsymbol{\gamma}|Y,X)  \propto l(\boldsymbol{\theta},\boldsymbol{\gamma}|Y,X) \times p(\boldsymbol{\theta}|\lambda) \times p(\boldsymbol{\gamma}|\tau) \times p(\lambda) \times p(\tau).
\end{align*}

As Gibbs sampling is not possible, we choose a Metropolis-Hastings sampling~\citep{hastings1970monte} to draw samples from the full conditional posterior and we provide an implementation in the R package \texttt{BDWreg}. From the posterior distribution, the mode of the marginal densities can be used as point estimate of the parameters, whereas the whole distribution is used for building credible intervals. In the case of Laplace priors, the inclusion or not of zero in the Highest Posterior Density (HPD) interval  is used for variable selection. MCMC samplers have been used before in the continuous Weibull regression context by \cite{newcombe2014weibull}, which utilizes a Reversible Jump MCMC, and \cite{soliman2012modified} which uses a hybrid method consisting of Metropolis-Hastings and Gibbs sampler to estimate parameters in a three parameters continuous Weibull distribution. Moreover,~\citep{polpo2009statistical} make use of a Metropolis-Hasting sampler to make inference for a continuous two-parameters Weibull distribution in a censoring framework.

\section{Simulations study}\label{DW:simulation}
In this section, we perform a simulation study where we show the effectiveness of the Bayesian estimation procedure, both in the case of data drawn from a DW regression model and in the case of model misspecification, where the generating model is that of Poisson or Negative Binomial (NB). Finally, we test the use of Laplace priors in a variable selection scenario.

\subsection{Simulation from a DW regression model}

Table (\ref{ch3:table1}) shows six configurations of parameters used in the simulation, where we consider the two link functions for $q$ and the link function for $\beta$ described in Section (\ref{sec:regression}), i.e. imposing a linear model on $\logit(q)$ or $\log(-\log(q))$, and on $\log(\beta)$.
\small
\begin{table}[!htp]
	\caption{The configuration of DW regression models used in the simulation.\label{ch3:table1}}
	\centering
\resizebox{\linewidth}{!}{
	\begin{tabular}{lll}
		\hline  \textbf{Model} & 	\multicolumn{2}{c}{\textbf{True Parameters}}\\
			\hline  $DW(q,\beta$) 		& $q =.41$	& $\beta=1.1$\\
			\hline  $DW(q,reg\beta$) 	&	$q=.8$	& 	 $\gamma_0 =.1$ , 	$\gamma_1=-.15$ , $\gamma_2=.5$	\\
			\hline  $\logit:DW(regQ,\beta$)		& $\theta_0 =.4$ , 	$\theta_1=-.1$ , $\theta_2=.34$	&  	$\beta=.7$\\
			\hline  $\logit:DW(regQ,reg\beta$)	& $\theta_0 =.4$ , 	$\theta_1=-.1$ , $\theta_2=.34$	&   	 $\gamma_0 =.1$ ,	 $\gamma_1=-.15$ , $\gamma_2=.5$  \\
			\hline  $\llog:DW(regQ,\beta$)		& $\theta_0 =.4$ , 	$\theta_1=-.1$ , $\theta_2=.34$	&  	$\beta=.7$  \\
			\hline  $\llog:DW(regQ,reg\beta$)	& $\theta_0 =.4$ , 	$\theta_1=-.1$ , $\theta_2=.34$	&   	 $\gamma_0 =.1$ ,	 $\gamma_1=-.15$ , $\gamma_2=.5$\\
			\hline		
	\end{tabular}}
\end{table}
\normalsize
For cases 2 to 6, we generate the three predictors uniformly in the interval $[0,1.5]$ and we simulate 500 observations. For the Bayesian estimation of the parameters, we use non-informative priors and make use of a Metropolis-Hastings algorithm with an independent Gaussian proposal to draw samples from the posterior. The scale of the proposal is adjusted so that a recommended acceptance rate lies in $(22,25)\%$~\citep{bedard2008optimal}. We consider 25,000 iterations of the sampler and use the first $25\%$ of the data as burn-in.

\begin{figure}[ht]
\centering
\includegraphics[width=\linewidth]{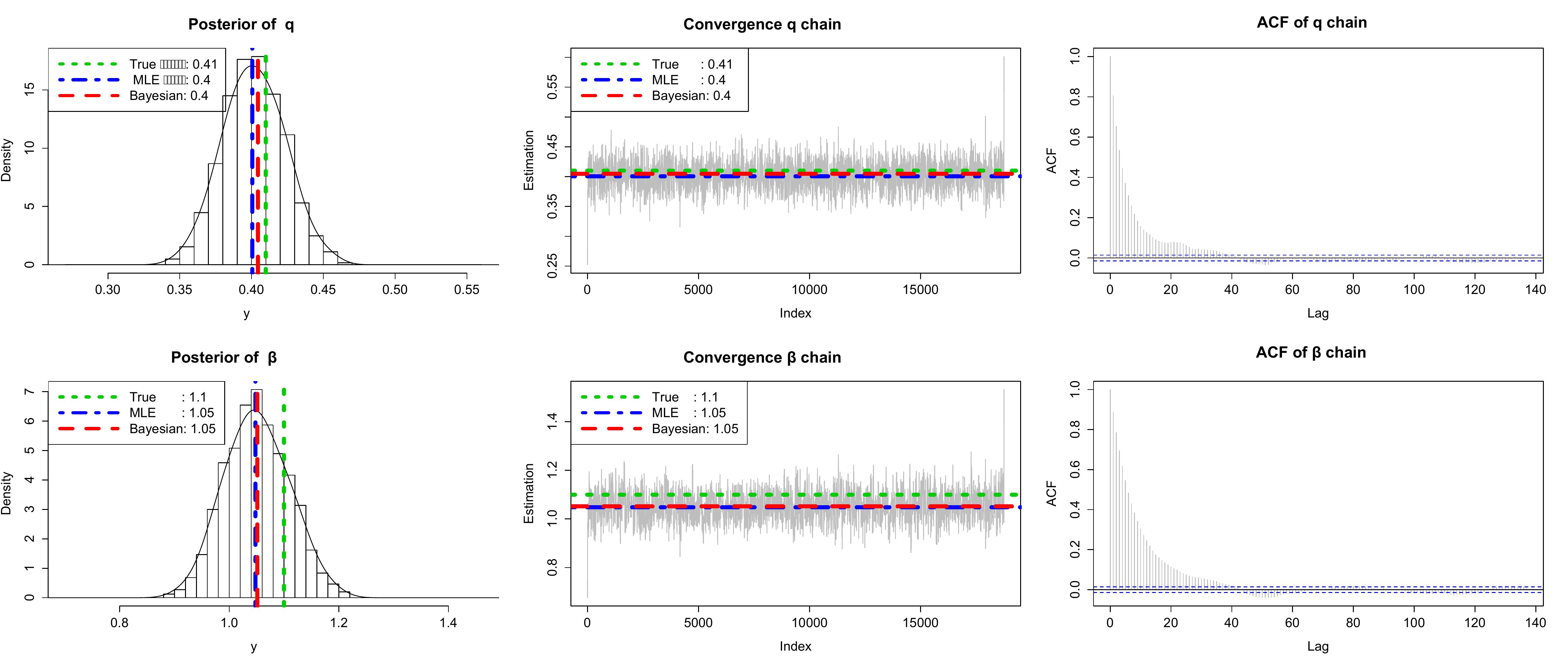}
\caption{Marginal densities and chain convergence for $q$ (top) and $\beta$ (bottom), for  case 1  where there are no exogenous variables in model.}
\label{fig:summary_case1}
\end{figure}

\begin{figure}[!hp]
\centering
\includegraphics[width=.75\linewidth]{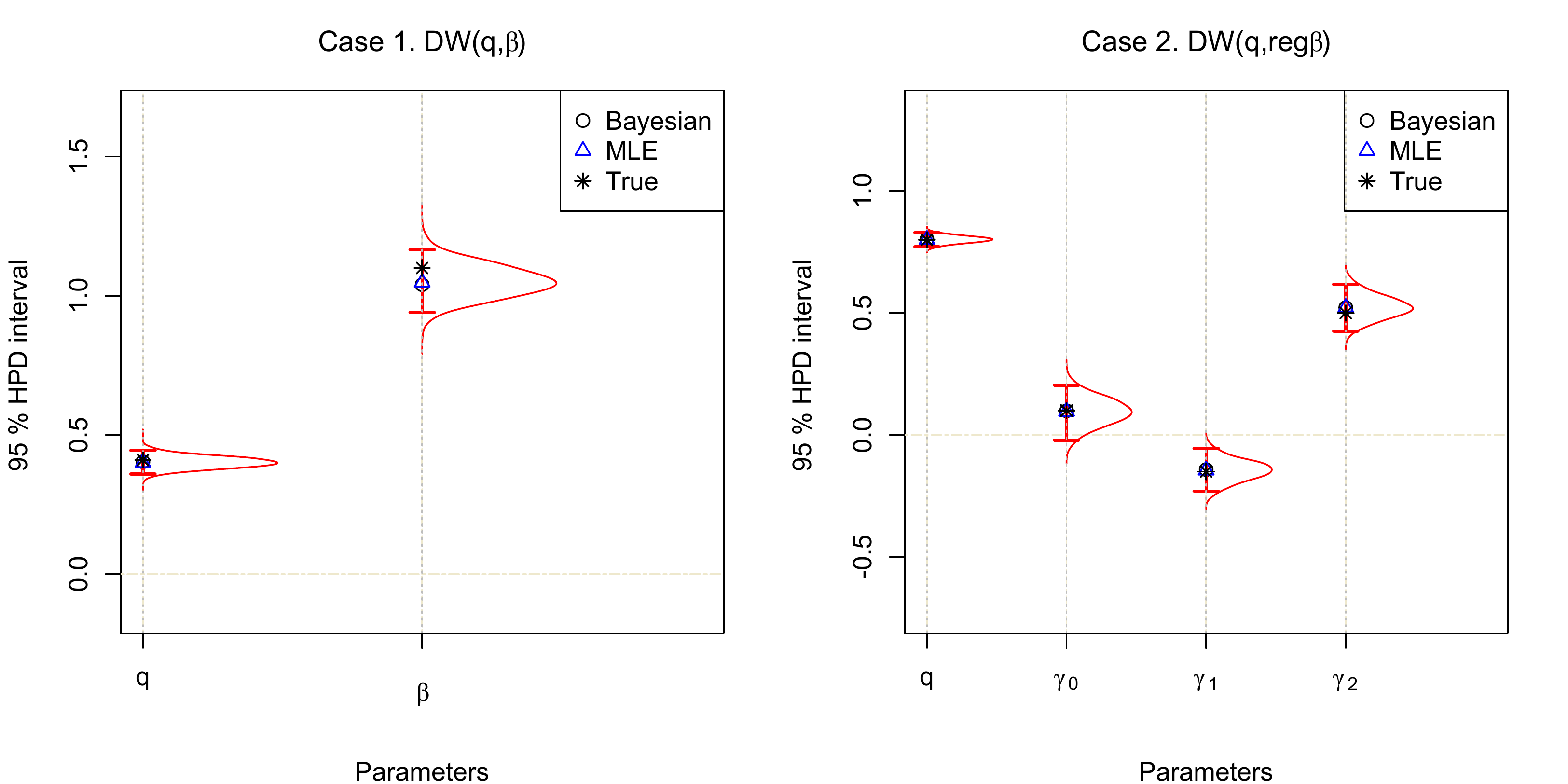}
\includegraphics[width=.75\linewidth]{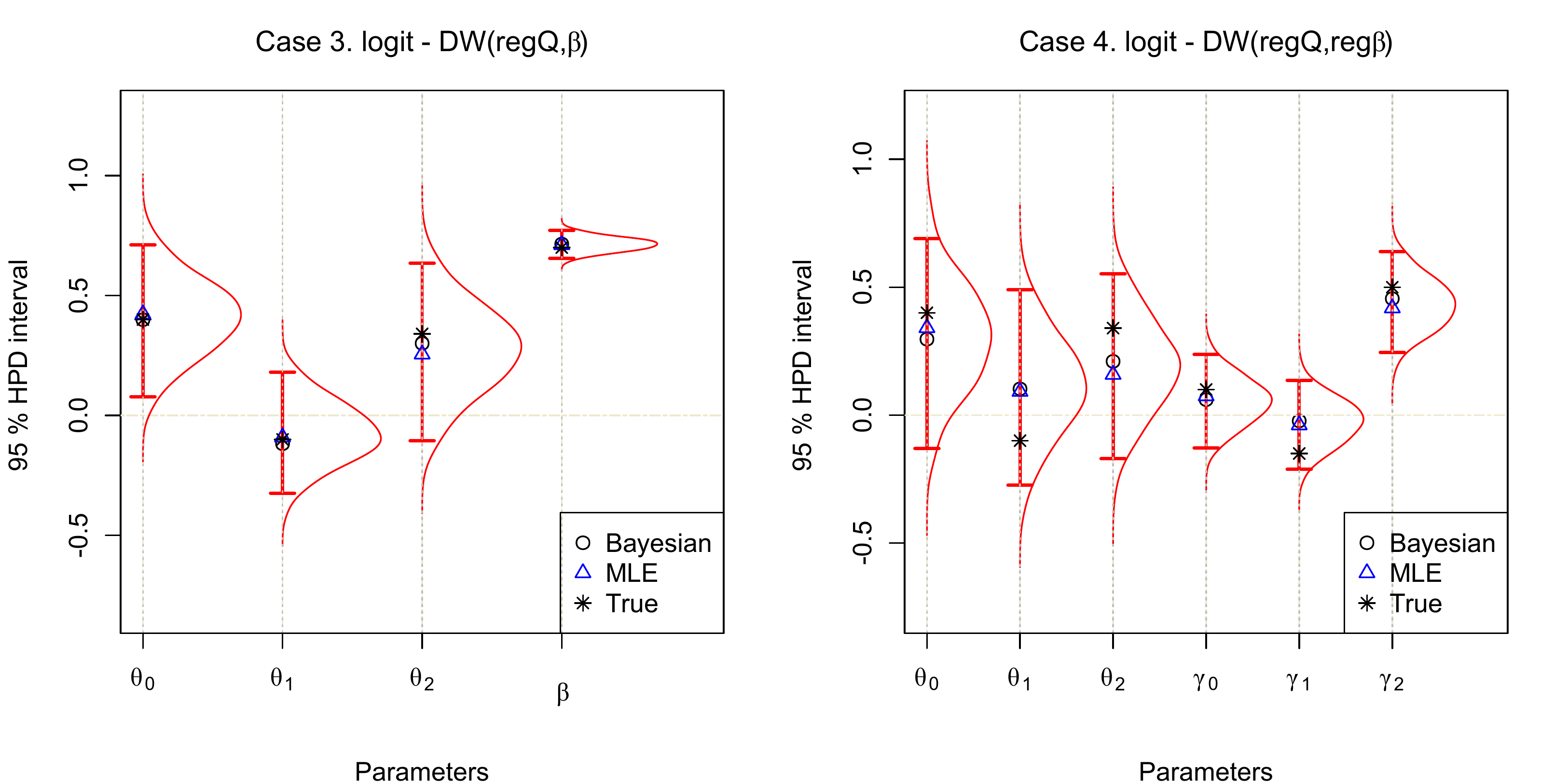}
\includegraphics[width=.75\linewidth]{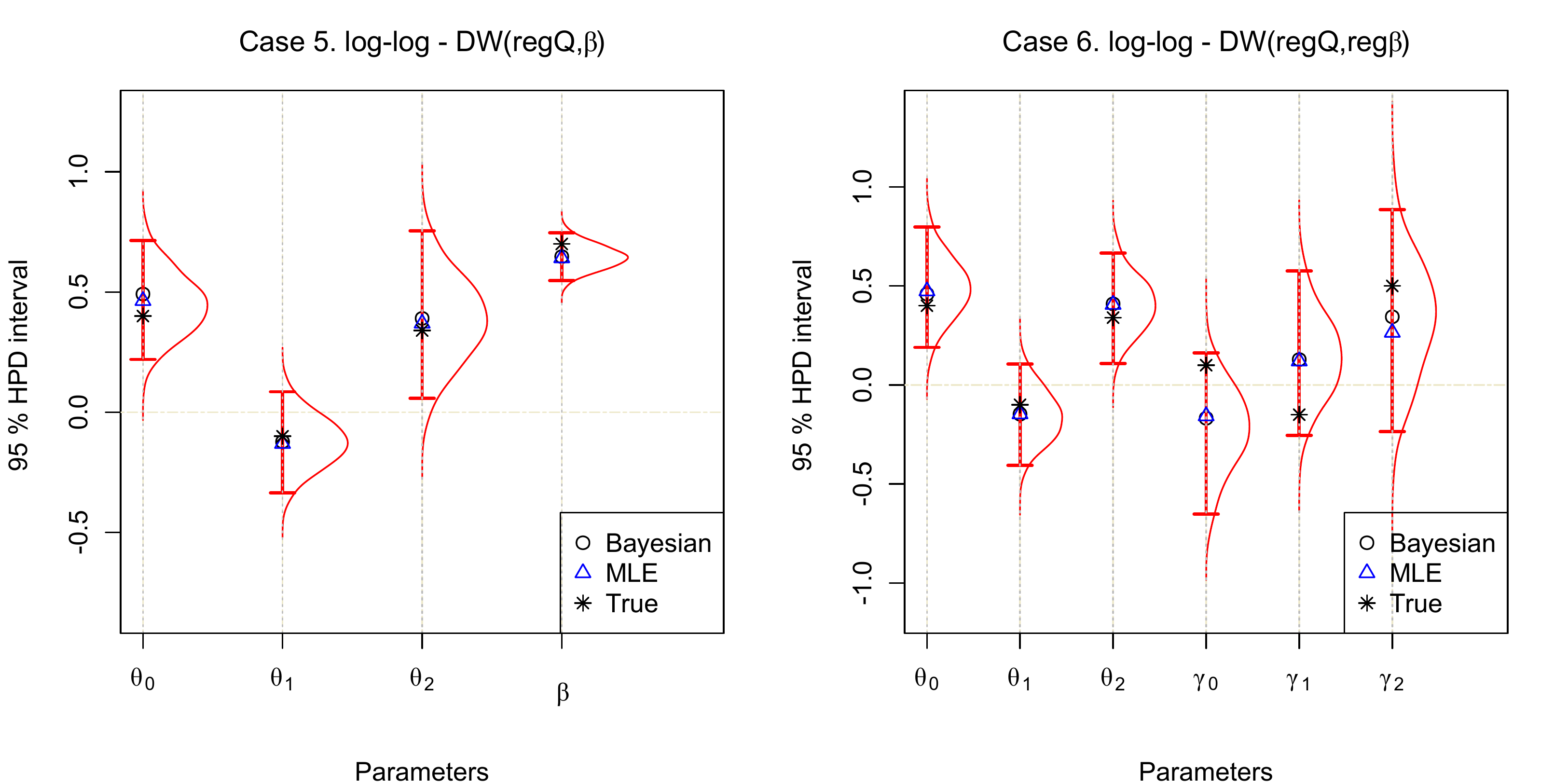}
\caption{Marginal densities and 95\% high probability density interval  for cases 1-6 in Table (\ref{ch3:table1}).}
\label{fig:case1234 HPDI}
\end{figure}

Figure (\ref{fig:summary_case1}) shows the posterior distribution of the parameters and the chain convergence in the first case, when no exogenous variables are present. Similar plots are obtained for the other cases. Figure (\ref{fig:case1234 HPDI}) shows the marginal densities of the parameters and the 95\% HPD interval for all six cases, as well as the maximum likelihood point estimate and the true value of the parameters. Overall, the plots show convergence of the chain and accurate estimation of the parameters.
\normalsize

\subsection{Simulation from a Poisson and NB regression model}

\begin{figure}[!hp]
\centering
\includegraphics[width=\linewidth]{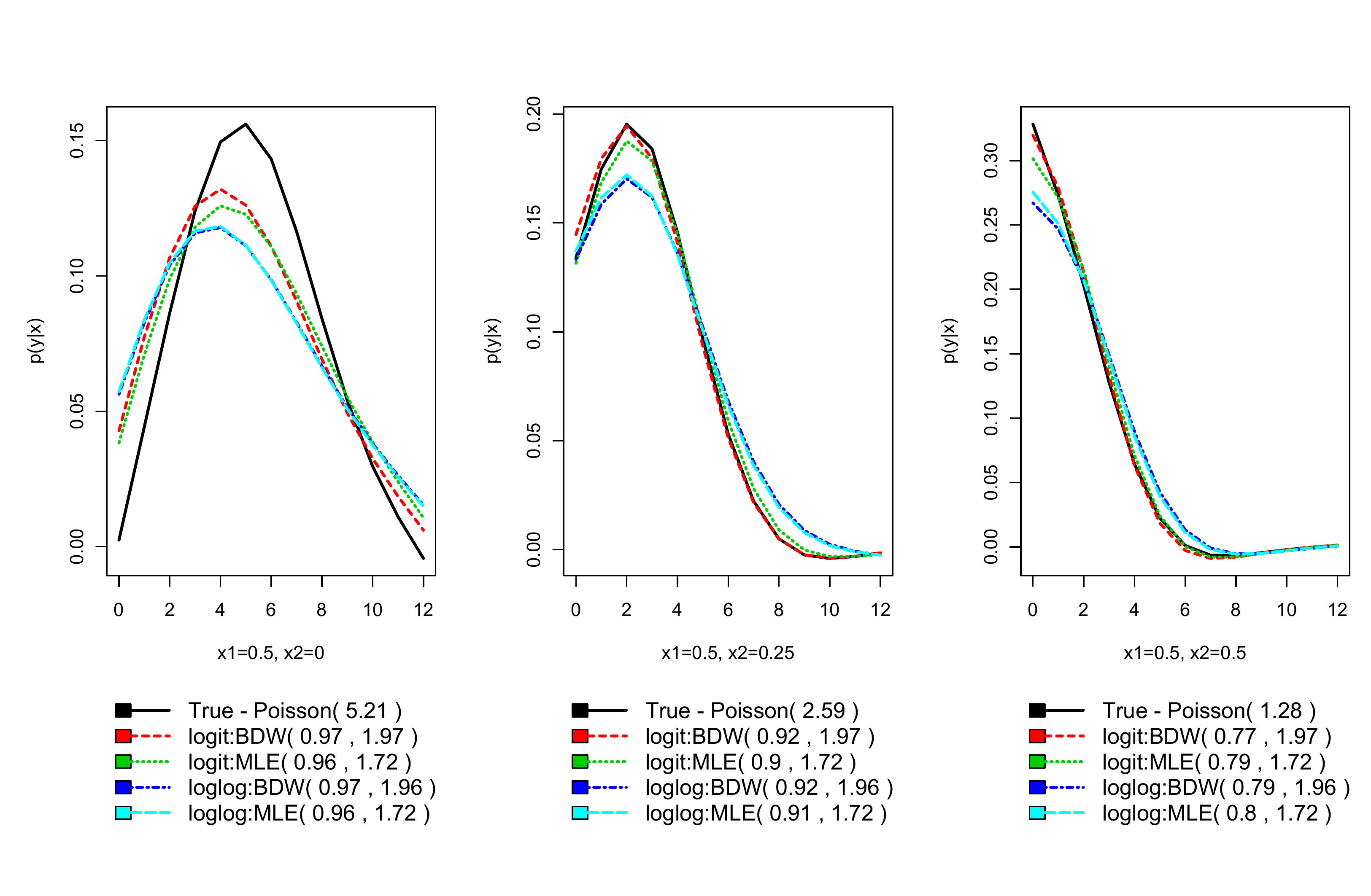}
\includegraphics[width=\linewidth]{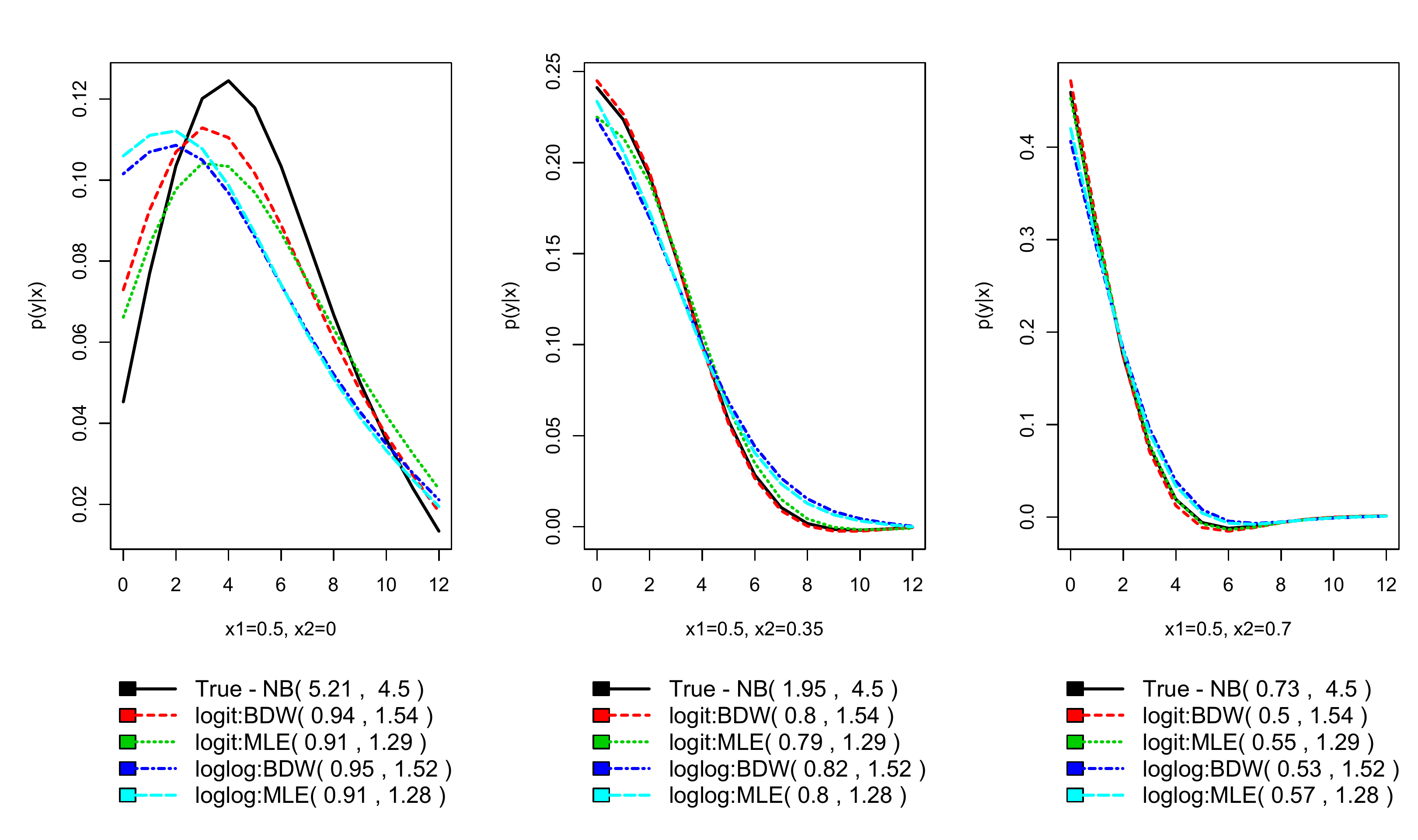}
\caption{Fitting Poisson (top) and NB (bottom) simulated data by $DW(regQ,\beta)$ for a range of values of $x_2$ and fixed $x_1 = 0.5$. The plots show the true conditional pmf (black) together with the conditional pmf fitted by the Bayesian DW  model proposed in this paper, with the $\logit(q)$ (red) and $\llog(q)$ (blue) links, and by the corresponding frequentist approaches (green and light blue, respectively).}
\label{fig:DW_SIM}
\end{figure}

The aim of this section is to test the fitting of a DW regression model to data generated from a Poisson and NB regression. To this end, we design two experiments using two explanatory variables, $X=(X_1,X_2)$, and $n=500$ data points. We simulate data for the predictors from uniform distributions, namely $X_1\sim U(0,1)$ and $X_2\sim U(0,1.5)$. We fix the intercept and the regression parameters to $\alpha=( -0.5,4.3,-2.2)$, with values chosen to cover a wide range of shapes for the target distribution. Then, in the first experiment, we assume that the conditional distribution of $Y$ given $X$ is ${\rm Poisson}(e^{X\alpha})$, whereas in the second experiment, we assume it to be a NB distribution with mean $\mu=e^{X\alpha}$ and variance $\mu + \mu^2/\theta$ with $\theta = 4.5$. Figures (\ref{fig:DW_SIM}) shows the conditional distribution fitted by $DW(regQ,\beta)$ for a fixed value of $x_1=0.5$ and sliding values of $x_2$ in the $[0,0.7]$ interval. 
 The figure shows how the estimation improves as the mean of the target distribution decreases, both for Bayesian and frequentist approaches. In addition, the $\logit$ link shows a better fit compared to the $\llog$ link in both Poisson and NB  experiments. For the frequentist estimation, we use the R package \texttt{DWreg}.

\subsection{Simulation on Variable Selection}
In this simulation, we show the performance of DW regression for variable selection. To this end, we generate $50$ parameters  uniformly in the $[-0.5,0.5]$ interval. Without loss of generality we assume that $75\%$ of the parameters, 37 out of 50,  are zero and the rest are non-zero. We simulate $500$ observations for each predictor from a $U(0,1.5)$ distribution, and the response variable from a DW distribution using a logit link for $q$ or the log link for $\beta$. Similar results are obtained with the $\llog$ link function. For parameter estimation, we keep the average rate of acceptance  in the $(20,30) \%$ interval for the total of $50,000$ iterations. Variable selection is performed by considering the 95\% HPD interval for each parameter.

Table (\ref{ch3:High-dimentional simulations}) shows the performance of the method in terms of selection of variables. In particular, the table reports the True Negative Rate (TNR), Recall $\Big(\dfrac{\rm{TP}}{\rm{TP}+\rm{FN}}\Big)$, Precision $\Big(\dfrac{\rm{TP}}{\rm{TP}+\rm{FP}}\Big)$ and $F_1$ score $\Big(\dfrac{2\rm{TP}}{2\rm{TP}+\rm{FN}+\rm{FP}}\Big)$, averaged over 20 simulations. The table shows a good performance overall, particularly for the $BDW(regQ,\beta$) models. The model with the $\log(\beta)$ link does not perform very well when $q$ decreases, i.e. when the number of zeros in the sample increases. In these cases, the models show a low recall, that is a high false negative rate.
\begin{table}[!hpt]
	\caption{Performance of BDW with Laplace priors. Variables are selected based on  the 95\% HPD interval and the selection is compared with the truth on the basis of True Negative Rate (TNR), recall, precision and $F_1$ score.\label{ch3:High-dimentional simulations}}
	\centering
	\resizebox{0.85\linewidth}{!}{%
	\begin{tabular}{lllccccccc}
	\hline \textbf{} & \textbf{Model}	&& \textbf{TNR} 	& \textbf{Recall} 	 & \textbf{Precision} &  \textbf{$\boldsymbol{F_1}$}\\
\hline &  $BDW(regQ,\beta=.1$) 	&    	&  93\%		 	&  90\%	&  	93\% & 91\%\\
	 & $BDW(regQ,\beta=.8$) 	 	&  		&  95\%		&  89\%	&  95\% & 92\% 	\\
	& $BDW(regQ,\beta=1.6$)  		&		&  93\%	&  91\%	&   93\% & 92\%	\\
\hline
	& $BDW(regQ,reg\beta$) 	&   		&  97\%	 &  68\%	& 96\% & 79\%	\\
\hline
	& $BDW(q=.85,reg\beta$) &   		&  90\%		 &  92\%	&  91\% & 91\%		\\
	 & $BDW(q=.50,reg\beta$) 	 	&   		&  93\%		 &  37\%		& 	84\% & 52\%	\\
	\hline
	\end{tabular}
	}
\end{table}

\section{Analysis of counts in medicine}\label{empirical data in medicine}

In this section, we show the performance of the Bayesian discrete Weibull regression model on real datasets from the medical domain. We compare the proposed model with the Bayesian Poisson (BPoisson) and Bayesian Negative Binomial (BNB) models on the basis of a number of commonly used criteria: Bayesian Information Criteria (BIC)~\citep{dayton2003model}, Akaike Information Criteria (AIC)~\citep{dayton2003model}, Deviance Information Criterion (DIC)~\citep{spiegelhalter2002bayesian}, Quasi-likelihood Information Criteria (QIC)~\citep{pan2001akaike}, Consistent AIC (CAIC)~\citep{bozdogan1987model}, Bayesian Predictive Information Criterion (BPIC)~\citep{ando2007bayesian} and  the Prior Predictive Density (PPD) used in the Bayes factor~\citep{kass1993bayes}.

\subsection{Comparison with Bayesian generalised linear models} \label{sec:Analysis of medical data}

To show the ability of BDW to estimate parameters in the presence of  under-dispersion, over-dispersion and excessive zeros in count data, we choose  the following three medical datasets:
\begin{enumerate}
\item The data on inhaler usage from \cite{grunwald2011statistical}, with 5209 observations. The response is the daily counts of inhalers usage, whereas the covariates are  humidity, barometric pressure, daily temperature, air particles level. The data show under-dispersion \citep{kalktawi15}.
\item The German health survey dataset available in the R package \texttt{COUNT} under the name \texttt{badhealth}, with 1127 observations. The response is the number of visits to doctors and the predictors are whether the patient claims to be in bad health or not, and the age of the patient. The data show over-dispersion \citep{kalktawi15}.
\item The German health registry dataset available in the R package \texttt{COUNT} under the name \texttt{rwm}, with 27326 observations. The response is the number of visits to doctors and the predictors are age, years of education and household yearly income. The data show over-dispersion with a relatively large number of zeros (37\%) \citep{kalktawi15}.
\end{enumerate}

We fit a BDW model with an uninformative prior on the regression parameters, 35000 iterations for the Metropolis-Hastings algorithm and an acceptance rate in the $(20,30)\%$ interval. For the case of BPoisson and BNB regression, we make use of  the \texttt{MCMCpack} R package~\citep{martin2011mcmcpack} with the same configurations. Table \ref{tb:RealData-Uninformative} shows a comparison of the models on the three datasets. We only report the results of the BDW(regQ,$\beta$) models and, of these, the \logit(q) link shows superior performance. In all cases, the BDW model has the same or better performance than both Poisson and negative Binomial. This was found also by \citep{kalktawi15}, where a frequentist approach was used and the comparison was made also with additional models such as COM-Poisson and hurdle/zero-inflated models.

\begin{table}[hp!]
	\centering
	\caption{Comparison of Bayesian DW, Poisson  and Negative Binomial on three datasets and under a number of information criteria. (*) denotes the minimum value.\label{tb:RealData-Uninformative}
	}
	\resizebox{\linewidth}{!}{%
	\begin{tabular}{lllllllll}
    \hline \textbf{Model} & \textbf{AIC} & \textbf{BIC}  & \textbf{CAIC} & \textbf{QIC} & \textbf{DIC} & \textbf{BPIC} & \textbf{log(PPD)} & \textbf{df}\\
	\hline \multicolumn{9}{l}{\textbf{Inhaler Use (under-dispersed)}}\\
	\hline  $\llog:BDW$&  13497.22  &  13536.57 & 13542.57 & 2.59* & 13487.63 & 13493.88 & -6745.93 & 6\\
	\hline  $\logit:BDW$& 13494.19*  &  13533.54* & 13539.54* & 2.59* & 13484.92* & 13490.49* & -6739.41* & 6\\
	\hline  BPoisson &   14009.01  &  14041.80 & 14046.80 & 2.69 & 13822.54 & 13734.31 & -6960.64 & 5\\
	\hline  BNB	   &  	 13952.85  &  13992.33 & 13998.20 & 2.68 & 13771.0 & 13686.47& -6960.81 & 6   \\
    \hline \multicolumn{9}{l}{\textbf{German Health Survey (over-dispersed)}}\\
  	\hline  $\llog:BDW$ & 	4478.9  &  4499.0 & 4502.0 & 3.98 & 4474.60 & 4478.33 & -2245.75 & 4\\
	\hline  $\logit:BDW$&   4475.2*	&  4495.3* & 4449.3* & 3.97* & 4474.16* & 4477.70* & -2242.23* & 4\\
	\hline  BPoisson	& 	5638.9	&  5654.02 & 5656.10 & 5.01 & 5638.14 & 5641.18 & -2826.88 & 3\\
	\hline  BNB		& 	4475.9	&  4495.9 & 4499.97 & 3.97* & 4474.66 & 4478.10 & -2243.87 & 4\\
    \hline \multicolumn{9}{l}{\textbf{German Health Registry (excessive zeros)}}\\
    \hline  $\llog:BDW$& 120340.1& 120381.2	& 120386.2	& 4.4* & 120334.6 & 120339.2 &  -60187.6 & 5\\
    \hline  $\logit:BDW$&  120339.2*& 120380.3*	& 120385.3*	&4.4*&  120327.0* & 120331.9* &  -60181.8* & 5\\
    \hline  BPoisson &  209636.4	& 209669.2	& 209673.2	& 7.7& 	209635.8 & 209639.6 & -104836.7 & 4\\
    \hline  BNB		&  120658.7	& 120708.0	& 120714.0	& 4.4*& 129125.8	 & 133365.3 & -60344.0  & 5\\
    \hline
	\end{tabular}
	}
	
 \end{table}

\subsection{Comparison with Bayesian penalised regression}

 In this section, we compare the performance of BDW to BPoisson and BNB regression for variable selection on a dataset with several variables. In particular, we consider the multivariate data of \cite{machado05}. The data consist of 5096 observations from the 1985 wave of the German Socioeconomic Panel. As in \cite{machado05}, we measure the demand in healthcare by the number of visits to a specialist (except gynecology or pedriatics) in the last quarter. The 20 covariates are listed in full in Table A.1 of \cite{machado05} and are the same considered in this paper. This is an extreme example of excessive zeros as the response variable contains 67.82\% of zeros.

 We fit a BDW model with  a Laplace prior on the regression parameters and a Gamma(2,1) hyper-prior on the shrinkage parameters. We consider 175000 iterations for the MCMC routine and similar configurations for the Bayesian Poisson and NB models. We also extend the comparison by including frequentist $L_1$ regularized models. In particular, we use the \texttt{glmnet} package~\citep{friedman2010regularization} to fit regularized Poisson regression and the \texttt{glm.nb} R function to fit regularized negative Binomial regression. In both cases, the penalty parameter is chosen by BIC. According to the results in Table (\ref{ch3:table3}), $DW(regQ,\beta)$ with the $\llog$ link achieves overall the best performance compared with the others BDW models and with NB and Poisson models.

\begin{table}[!hpt]
	\caption{Comparison of BDW with Bayesian and regularized NB and Poisson on the number of visits to a specialist dataset of \cite{machado05}. ($*$) denotes the minimum value, whereas df is the number of non-zero coefficients. For the Bayesian models, these are based on the 95\% HPD interval.\label{ch3:table3}}
	\centering
	\resizebox{\linewidth}{!}{%
	\begin{tabular}{clcccccccc}
		 \hline	& \textbf{Model}& \textbf{AIC} & \textbf{BIC} &	\textbf{CAIC} &  \textbf{QIC} & \textbf{DIC}  & \textbf{BPIC} 	 & \textbf{log(PPD)}  &	 \textbf{df}$$\\
		\hline  		& \logit:BDW(regQ,$\beta$)		& 12720.4	 & 	12864.2 & 	2.5* & 	12886.2 & 	12710.8	 & 12731.5 & 	-6392.3 & 	11\\
		\hline  		& \llog:BDW(regQ,$\beta$)		& 12698.5*	 & 	 12842.3* & 	 2.5* & 	12864.3* & 	12693.3* & 	12713.6* & 	-6383.3* & 	11\\
		\hline  		& BDW(q,reg$\beta$)		& 13256.0	 & 	13399.8 & 	2.6	 & 	13421.8	 & 	13250.4	 & 	13270.3	 & 	-6665.8	 & 	6 \\
		\hline  		& BPoisson& 21588.2	 & 	21705.8	 & 	 4.2 & 21723.8 &  	21594.6	 & 	21615.8		 & 	-10832.6		 & 	17\\
		\hline  		& BNB& 12867.3	 & 	12939.2	 & 2.5* &12950.2  & 	12838.3	 & 	12834.8		 & 	-6452.3		 & 11\\
		\hline  		& Poisson (glmnet)& 21571.1	 & 	21706.1	 & 	4.2 & 	21724.1	 & 	-		 & 	-		 & -		 & 	17\\
		\hline  		& NB (glm.nb)& 12839.3	 & 	12911.2	 & 	2.5* & 	12922.6	 & 	-		 & 	-		 & -		 & 	12\\
		\hline
			 	
	\end{tabular} %
	}
\end{table}
\normalsize

\begin{figure}[hpt!]
	\centering
	\includegraphics[width=\linewidth ]{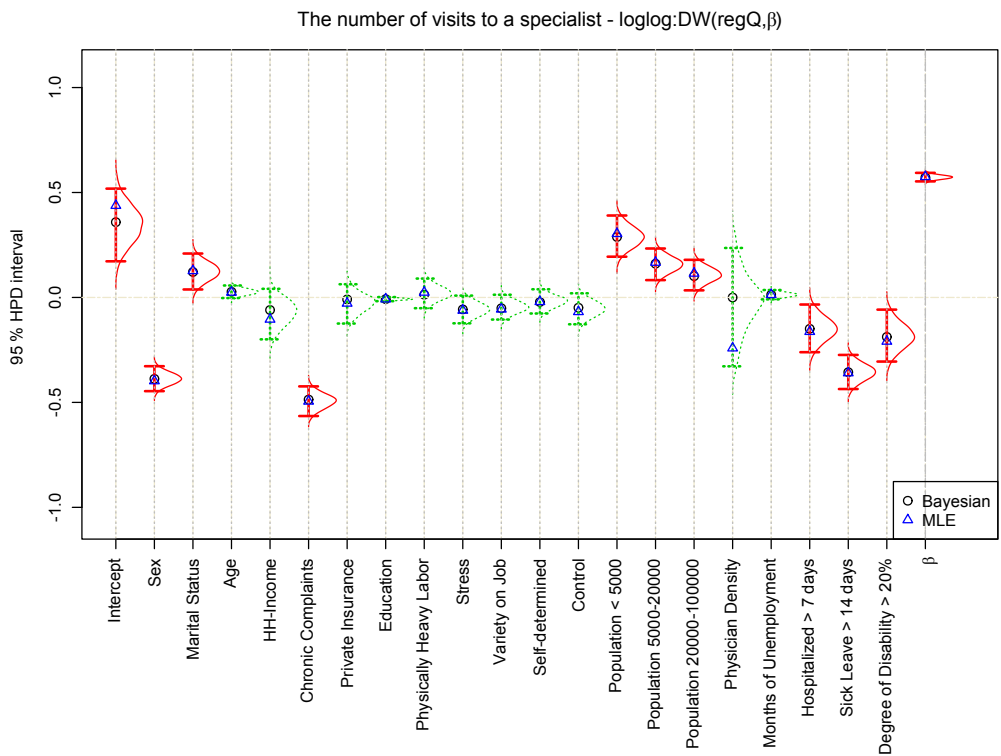}
	\caption{Marginal densities of the parameters for the  $BDW(regQ,\beta)$ model with \llog(q) link on the number of visits to a specialist dataset. The red lines are for the cases where the $95\%$ HDP interval does not contain zero (significant variable). Green dotted lines for the opposite.}
	\label{fig:Visiting_specialist_QregVsBreg}
\end{figure}

\begin{figure}[tph!]
\centering
\includegraphics[width=0.8\linewidth]{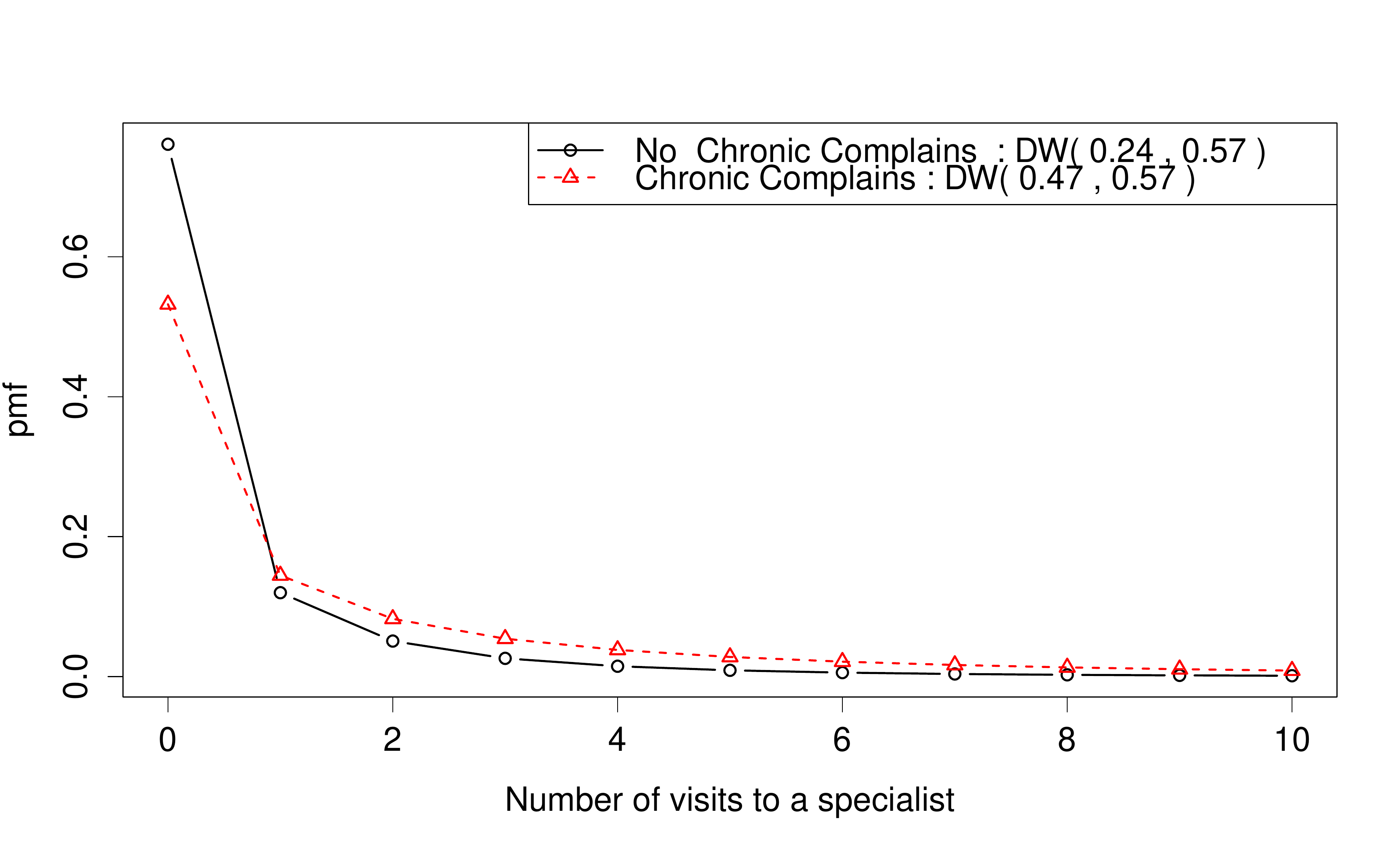}
\caption{Effect of the variable Chronic Complaints on the conditional distribution for
the healthcare data, when all other variables are held constant.}
\label{fig:CC_density_plot}
\end{figure}

Figure (\ref{fig:Visiting_specialist_QregVsBreg}) shows the marginal densities of the parameters for the $DW(regQ,\beta)$ with the $\llog$ link. Highlighted in red are those variables that are found to be significant based on the 95\% HPD interval. The selection is overall in accordance with the results obtained by \cite{machado05} using a jittering approach, with variables such as gender, chronic complaints, sick leave and disability found to be significant, and other variables like unemployment, private insurance and those related to job characteristics, such as heavy labor, stress, variety on job, self-determined and control found not to be significant. Figure \ref{fig:CC_density_plot} shows the effect of the variable chronic complaints on the conditional distribution, suggesting that the probability of a large number of visits is higher for the case of chronic complaints than for the case of no complaints. Table (\ref{significant_variables_in_TNVTS}) further compares the selection of variables with those selected by Poisson and NB regression models. Overall, there is high agreement between BDW and NB, with the exception of the variable control which is found significant by NB (both in the Bayesian and frequentist estimation) but not by DW. Poisson and BPoisson tend to select many more variables.

\begin{table}[!ht]
	\caption{Significant covariates that are selected by $BDW(regQ,\beta)$ with \llog link, Bayesian and regularized $NB$ and $Poisson$ regression models for the number of visits to a specialist dataset. An $(*)$ indicates a non-zero coefficient.}
	\centering
	\resizebox{\linewidth}{!}{%
	\begin{tabular}{lccccc}
	\textbf{Variable} 	& \textbf{BDW(regQ,$\beta$)}& \textbf{NB} 	&  \textbf{BNB} & \textbf{Poisson}  & \textbf{BPoisson}\\
	\hline
	 	 Sex						 & 	*		&		*		&	*	& * &*\\ 
	 Marital status	 				 & 	*		&		*		&	*	& * &*\\ 
	 Age	 						 & 	 			&				&		&* & \\ 
	 HH-income						 & 	 			&				&		&*& *\\ 
	 Chronic complaints				 & 	*		&		*		&	*	&* & * \\ 
	Private insurance	 			 & 	  			&				&		& & \\ 
	 Education	 				 	 & 	 			&				&		&*&\\ 
	 Physically heavy labour		 & 	 			&				&		&*&*\\ 
	 Stress	 						 & 	 			&				&		&*&* \\ 
	 	 Variety on job	 			  	 & 	 			&				&		&*&*\\ 
		 Self-determined				 & 	 			&				&		&&\\ 
		 Control	 				 	 & 	 			&		*		&	*	&*&*\\ 
	 	 Population < 5000				 & 	* 		&		*		&	*	&*&*\\ 
	 Population 5000-20000			 & 	* 		&		*		&	*	&*&*\\ 
	 	 Population 20000-100000		 & 	* 		&		*		&	*	&*&*\\ 
	 	 Physician density	 			 & 	 			&				&		&& *\\ 
	 	 Months of unemployment			 & 	 			&				&	*	&&*\\ 
		 Hospitalized > 7 days			 & 	* 		&		*		&	*	&*&*\\ 
		 Sick Leave > 14 days	 		 & 	* 		&		*		&	*	&*&*\\ 
	 	 Degree of disability > 20	  	 & 	* 		&		*		&	 *	&*&*\\ 
	\hline
	\end{tabular} %
	}
	\label{significant_variables_in_TNVTS}
\end{table}
\normalsize

\section{Conclusion}\label{Conclusion and discussion}
In this paper we have proposed a novel Bayesian regression model for count data, by assuming a discrete Weibull conditional distribution. We have shown the applicability of this method to count data from the medical domain. In particular, we analyse datasets on the number of visits to doctors/specialists, a quantity that is often used as an indicator of healthcare demand. The response variable in the examples considered is discrete and is characterized by a skewed distribution, making the whole conditional distribution of interest and not only the conditional mean.

We have tested the inference procedure on simulated and real data with various characteristics, such as under-dispersion, over-dispersion and excess of zeros.  Overall, we have found a good performance of the method in comparison with Poisson and NB regression models, on the basis of a number of information criteria and of the selection of influential variables. The method is implemented in the R package \texttt{BDWreg}, which is available in CRAN. Future work will explore an extension of the approach proposed in this paper to more flexible DW regression models, such as zero-inflated, multilevel and mixture DW models, in a similar spirit to the existing models for continuous responses \citep{dunson07}.

\bibliographystyle{elsarticle-harv} 
\bibliography{reference}

\begin{thebibliography}{45}
\expandafter\ifx\csname natexlab\endcsname\relax\def\natexlab#1{#1}\fi
\expandafter\ifx\csname url\endcsname\relax
  \def\url#1{\texttt{#1}}\fi
\expandafter\ifx\csname urlprefix\endcsname\relax\def\urlprefix{URL }\fi

\bibitem[{Ando(2007)}]{ando2007bayesian}
Ando, T., 2007. {Bayesian} predictive information criterion for the evaluation
  of hierarchical {Bayesian} and empirical {Bayes} models. Biometrika 94~(2),
  443--458.

\bibitem[{Angers and Biswas(2003)}]{angers2003bayesian}
Angers, J.-F., Biswas, A., 2003. A {Bayesian} analysis of zero-inflated
  generalized {Poisson} model. Computational Statistics and Data Analysis
  42~(1), 37--46.

\bibitem[{Ara\'ujo~Santos and Fraga~Alves(2013)}]{santos13}
Ara\'ujo~Santos, P., Fraga~Alves, M.~I., 2013. Improved shape parameter
  estimation in a discrete {Weibull} model. In: Recent Developments in Modeling
  and Applications in Statistics . Studies in Theoretical and Applied
  Statistics. Springer-Verlag, pp. 71--80.

\bibitem[{Atienza et~al.(2008)Atienza, Garcia-Heras, Munoz-Pichardo, and
  Villa}]{atienza08}
Atienza, N., Garcia-Heras, J., Munoz-Pichardo, J.~M., Villa, R., 2008. An
  application of mixture distributions in modelization of length of hospital
  stay. Statistics in Medicine 27~(9), 1403--1420.

\bibitem[{Bao et~al.(2014)Bao, Vinciotti, Wit, and 't~Hoen}]{bao14}
Bao, Y., Vinciotti, V., Wit, E., 't~Hoen, P., 2014. Joint modeling of
  {ChIP-seq} data via a {Markov} random field model. Biostatistics 15~(2),
  296--310.

\bibitem[{Bedard(2008)}]{bedard2008optimal}
Bedard, M., 2008. Optimal acceptance rates for metropolis algorithms: Moving
  beyond 0.234. Stochastic Processes and their Applications 118~(12),
  2198--2222.

\bibitem[{Bozdogan(1987)}]{bozdogan1987model}
Bozdogan, H., 1987. Model selection and {Akaike's information criterion (AIC)}:
  The general theory and its analytical extensions. Psychometrika 52~(3),
  345--370.

\bibitem[{Bracquemond and Gaudoin(2003)}]{bracquemond03}
Bracquemond, C., Gaudoin, O., 2003. A survey on discrete lifetime
  distributions. International Journal of Reliability, Quality and Safety
  Engineering 10~(01), 69--98.

\bibitem[{Cameron and Trivedi(2013)}]{cameron2013regression}
Cameron, A.~C., Trivedi, P.~K., 2013. Regression analysis of count data.
  Cambridge university press.

\bibitem[{Carter and Potts(2014)}]{carter14}
Carter, E., Potts, H., 2014. Predicting length of stay from an electronic
  patient record system: a primary total knee replacement example. BMC Medical
  Informatics and Decision Making 14~(26).

\bibitem[{Dayton(2003)}]{dayton2003model}
Dayton, C.~M., 2003. Model comparisons using information measures. Journal of
  Modern Applied Statistical Methods 2~(2), 2.

\bibitem[{Dunson et~al.(2007)Dunson, Pillai, and Park}]{dunson07}
Dunson, D.~B., Pillai, N., Park, J.-H., 2007. Bayesian density regression.
  Journal of the Royal Statistical Society: Series B (Statistical Methodology)
  69~(2), 163--183.

\bibitem[{El-Sayyad(1973)}]{el1973bayesian}
El-Sayyad, G., 1973. {Bayesian} and classical analysis of {Poisson} regression.
  Journal of the Royal Statistical Society. Series B (Methodological),
  445--451.

\bibitem[{Englehardt and Li(2011)}]{englehardt11}
Englehardt, J.~D., Li, R., 2011. The discrete {Weibull} distribution: An
  alternative for correlated counts with confirmation for microbial counts in
  water distributions. Risk Analysis 31~(3), 370--381.

\bibitem[{Esnaola et~al.(2013)Esnaola, Puig, Gonzalez, Castelo, and
  Gonzalez}]{esnaola13}
Esnaola, M., Puig, P., Gonzalez, D., Castelo, R., Gonzalez, J.~R., 2013. A
  flexible count data model to fit the wide diversity of expression profiles
  arising from extensively replicated {RNA}-seq experiments. BMC Bioinformatics
  14, 254.

\bibitem[{Friedman et~al.(2010)Friedman, Hastie, and
  Tibshirani}]{friedman2010regularization}
Friedman, J., Hastie, T., Tibshirani, R., 2010. Regularization paths for
  generalized linear models via coordinate descent. Journal of Statistical
  Software 33~(1), 1.

\bibitem[{Ghosh et~al.(2006)Ghosh, Mukhopadhyay, and Lu}]{ghosh2006bayesian}
Ghosh, S.~K., Mukhopadhyay, P., Lu, J.-C.~J., 2006. {Bayesian} analysis of
  zero-inflated regression models. Journal of Statistical planning and
  Inference 136~(4), 1360--1375.

\bibitem[{Grunwald et~al.(2011)Grunwald, Bruce, Jiang, Strand, and
  Rabinovitch}]{grunwald2011statistical}
Grunwald, G.~K., Bruce, S.~L., Jiang, L., Strand, M., Rabinovitch, N., 2011. A
  statistical model for under-or overdispersed clustered and longitudinal count
  data. Biometrical Journal 53~(4), 578--594.

\bibitem[{Hastings(1970)}]{hastings1970monte}
Hastings, W.~K., 1970. {Monte} {Carlo} sampling methods using {Markov} chains
  and their applications. Biometrika 57~(1), 97--109.

\bibitem[{Hougaard et~al.(1997)Hougaard, Lee, and Whitmore}]{hougaard97}
Hougaard, P., Lee, M.~T., Whitmore, G.~A., 1997. Analysis of overdispersed
  count data by mixtures of {Poisson} variables and {Poisson} processes.
  Biometrics 53, 1225--1238.

\bibitem[{Ishwaran and Rao(2005)}]{ishwaran2005spike}
Ishwaran, H., Rao, J.~S., 2005. Spike and slab variable selection: frequentist
  and {Bayesian} strategies. Annals of Statistics, 730--773.

\bibitem[{{Kalktawi} et~al.(2015){Kalktawi}, {Vinciotti}, and
  {Yu}}]{kalktawi15}
{Kalktawi}, H.~S., {Vinciotti}, V., {Yu}, K., Nov. 2015. {A Simple and Adaptive
  Dispersion Regression Model for Count Data}. arXiv:1511.00634v1.

\bibitem[{Kass(1993)}]{kass1993bayes}
Kass, R.~E., 1993. {Bayes} factors in practice. The Statistician, 551--560.

\bibitem[{Khan et~al.(1989)Khan, Khalique, and Abouammoth}]{khan89}
Khan, M. S.~A., Khalique, A., Abouammoth, A.~M., 1989. On estimating parameters
  in a discrete {Weibull} distribution. IEEE transactions on Reliability
  38~(3), 348--350.

\bibitem[{Kulasekera(1994)}]{kulasekera94}
Kulasekera, K.~B., 1994. Approximate {MLE}'s of the parameters of a discrete
  {Weibull} distribution with type {I} censored data. Microelectronics
  Reliability 34~(7), 1185--1188.

\bibitem[{Kyung et~al.(2010)Kyung, Gill, Ghosh, Casella,
  et~al.}]{kyung2010penalized}
Kyung, M., Gill, J., Ghosh, M., Casella, G., et~al., 2010. Penalized
  regression, standard errors, and {Bayesian} lassos. {Bayesian} Analysis
  5~(2), 369--411.

\bibitem[{Lai(2013)}]{lai13}
Lai, C.~D., 2013. Issues concerning constructions of discrete lifetime models.
  Qualitative technology of quantitative managment 10~(2), 251--262.

\bibitem[{Lam et~al.(2006)Lam, Xue, and Bun~Cheung}]{lam06}
Lam, K.~F., Xue, H., Bun~Cheung, Y., 2006. Semiparametric analysis of
  zero-inflated count data. Biometrics 62~(4), 996--1003.

\bibitem[{Liu and Powers(2012)}]{liu2012bayesian}
Liu, H., Powers, D.~A., 2012. {Bayesian} inference for zero-inflated {Poisson}
  regression models. Journal of Statistics: Advances in Theory and Applications
  7~(2), 155--188.

\bibitem[{Machado and Santos~Silva(2005)}]{machado05}
Machado, J., Santos~Silva, M., 2005. Quantiles for counts. JASA 100~(472),
  1226--1237.

\bibitem[{Martin et~al.(2011)Martin, Quinn, Park, et~al.}]{martin2011mcmcpack}
Martin, A.~D., Quinn, K.~M., Park, J.~H., et~al., 2011. {MCMCpack}: {Markov}
  chain {Monte Carlo} in {R}. Journal of Statistical Software 42~(9), 1--21.

\bibitem[{Mohebbi et~al.(2014)Mohebbi, Wolfe, and Forbes}]{mohebbi2014disease}
Mohebbi, M., Wolfe, R., Forbes, A., 2014. Disease mapping and regression with
  count data in the presence of overdispersion and spatial autocorrelation: a
  {Bayesian} model averaging approach. International journal of environmental
  research and public health 11~(1), 883--902.

\bibitem[{Nagakawa and Osaki(1975)}]{nakagawa75}
Nagakawa, T., Osaki, S., 1975. The discrete {Weibull} distribution. IEEE
  transactions on reliability R-24~(5).

\bibitem[{Neelon et~al.(2010)Neelon, O'Malley, and
  Normand}]{neelon2010bayesian}
Neelon, B.~H., O'Malley, A.~J., Normand, S.-L.~T., 2010. A {Bayesian} model for
  repeated measures zero-inflated count data with application to outpatient
  psychiatric service use. Statistical Modelling 10~(4), 421--439.

\bibitem[{Newcombe et~al.(2014)Newcombe, Ali, Blows, Provenzano, Pharoah,
  Caldas, and Richardson}]{newcombe2014weibull}
Newcombe, P., Ali, H.~R., Blows, F., Provenzano, E., Pharoah, P., Caldas, C.,
  Richardson, S., 2014. {Weibull} regression with {Bayesian} variable selection
  to identify prognostic tumour markers of breast cancer survival. Statistical
  methods in medical research, 0962280214548748.

\bibitem[{Ozsolak and Milos(2011)}]{ozsolak11}
Ozsolak, F., Milos, P.~M., 2011. {RNA} sequencing: advances, challenges and
  opportunities. Nature Review Genetics 12, 87--98.

\bibitem[{Pan(2001)}]{pan2001akaike}
Pan, W., 2001. Akaike's information criterion in generalized estimating
  equations. Biometrics 57~(1), 120--125.

\bibitem[{Park and Casella(2008)}]{park2008bayesian}
Park, T., Casella, G., 2008. The {Bayesian} lasso. Journal of the American
  Statistical Association 103~(482), 681--686.

\bibitem[{Polpo et~al.(2009)Polpo, Coque~Jr, and
  Pereira}]{polpo2009statistical}
Polpo, A., Coque~Jr, M., Pereira, C., 2009. Statistical analysis for {Weibull}
  distributions in presence of right and left censoring. In: Reliability,
  Maintainability and Safety, 2009. ICRMS 2009. 8th International Conference
  on. IEEE, pp. 219--223.

\bibitem[{Robinson and Smyth(2008)}]{robinson08}
Robinson, M.~D., Smyth, G.~K., 2008. Small-sample estimation of negative
  binomial dispersion, with applications to {SAGE} data. Biostatistics 9~(2),
  321--332.

\bibitem[{Sellers and Shmueli(2010)}]{sellers10}
Sellers, K.~F., Shmueli, G., 2010. A flexible regression model for count data.
  The Annals of Applied Statistics 4~(2), 943--961.

\bibitem[{Soliman et~al.(2012)Soliman, Abd-Ellah, Abou-Elheggag, and
  Ahmed}]{soliman2012modified}
Soliman, A.~A., Abd-Ellah, A.~H., Abou-Elheggag, N.~A., Ahmed, E.~A., 2012.
  Modified {Weibull} model: A {Bayes} study using {MCMC} approach based on
  progressive censoring data. Reliability Engineering \& System Safety 100,
  48--57.

\bibitem[{Spiegelhalter et~al.(2002)Spiegelhalter, Best, Carlin, and Van
  Der~Linde}]{spiegelhalter2002bayesian}
Spiegelhalter, D.~J., Best, N.~G., Carlin, B.~P., Van Der~Linde, A., 2002.
  {Bayesian} measures of model complexity and fit. Journal of the Royal
  Statistical Society: Series B (Statistical Methodology) 64~(4), 583--639.

\bibitem[{Tibshirani(1996)}]{tibshirani1996regression}
Tibshirani, R., 1996. Regression shrinkage and selection via the lasso. Journal
  of the Royal Statistical Society. Series B (Methodological), 267--288.

\bibitem[{Zhou et~al.(2012)Zhou, Li, Dunson, and Carin}]{zhou12}
Zhou, M., Li, L., Dunson, D., Carin, L., 2012. {Lognormal and gamma mixed
  negative binomial regression}. In: Proceedings of the 29th International
  Conference on Machine Learning. Vol. 2012. NIH Public Access, p. 1343.

\end{thebibliography}
\end{document}